\documentclass[english,aps,twocolumn,showpacs,prl,superscriptaddress,floatfix]{revtex4-1}

\usepackage{bm}
\usepackage{hyperref}
\usepackage{amsmath}
\usepackage{amssymb}
\usepackage{xspace}
\usepackage{graphicx}
\usepackage{bbm}
\usepackage{bbold}

\newcommand{\ie}{i.e.}

\newcommand{\ai}{\textit{ab initio}\xspace}
\newcommand{\Acm}{\ensuremath{\mathrm{A}\mathrm{cm}^{-2}}\xspace}  
\newcommand{\sdl}{\ensuremath{l_{s}}\xspace}

\usepackage{booktabs} 

\begin{document}

\title{Magneto-Optical Detection of the Spin Hall Effect in Pt and W Thin Films}

\author{C. Stamm} 
\author{C. Murer}
\affiliation{Department of Materials, ETH Z{\"u}rich, 8093 Z{\"u}rich, Switzerland}
\author{M. Berritta} 
\affiliation{Department of Physics and Astronomy, Uppsala University, P.\,O.\ Box 516, SE-75120 Uppsala, Sweden}
\author{J. Feng}
\affiliation{Department of Materials, ETH Z{\"u}rich, 8093 Z{\"u}rich, Switzerland}
\author{M. Gabureac}
\affiliation{Department of Materials, ETH Z{\"u}rich, 8093 Z{\"u}rich, Switzerland}
\author{P. M. Oppeneer}
\affiliation{Department of Physics and Astronomy, Uppsala University, P.\,O.\ Box 516, SE-75120 Uppsala, Sweden}
\author{P. Gambardella}
\affiliation{Department of Materials, ETH Z{\"u}rich, 8093 Z{\"u}rich, Switzerland}


\begin{abstract}
The conversion of charge currents into spin currents in nonmagnetic conductors is a hallmark manifestation of spin–-orbit coupling that has important implications for spintronic devices. Here we report the measurement of the interfacial spin accumulation induced by the spin Hall effect in Pt and W thin films using magneto-optical Kerr microscopy. We show that the Kerr rotation has opposite sign in Pt and W and scales linearly with current density.
By comparing the experimental results with \ai calculations of the spin Hall and magneto-optical Kerr effects, we quantitatively determine the current-induced spin accumulation at the Pt interface as $5 \cdot 10^{-12}$~$\mu_B$A$^{-1}$cm$^2$ per atom. From thickness-dependent measurements, we determine the spin diffusion length in a single Pt film to be $11 \pm 3$~nm, which is significantly larger compared to that of Pt adjacent to a magnetic layer.
\end{abstract}


\maketitle

The spin Hall effect (SHE) converts an electric charge current flowing along a wire into a transverse spin current \cite{Dyakonov1971,Hirsch1999,Sinova2015}, leading to the accumulation of spins at the surface of the wire \cite{Zhang2000}. In nonmagnetic metals (NM), the induced spin polarization is usually detected \textit{indirectly} through its interaction with an adjacent ferromagnet (FM). Experimental methods to measure the SHE rely on the nonlocal resistance in lateral spin valve devices, in which the NM is either in contact \cite{Valenzuela2006} or separated from the FM electrodes \cite{Niimi2015}, as well as on the spin pumping effect \cite{Saitoh2006}, the spin Hall magnetoresistance \cite{Nakayama2013,Avci2015}, and the detection of the SHE-induced spin-orbit torques \cite{Ando2008,Liu2011,Garello2013,Fan2014} and magnetization reversal \cite{Miron2011,Liu2012a} in NM/FM bilayers. In such systems, however, magnetization-dependent scattering, interfacial spin-orbit coupling, and proximity effects deeply influence the spin accumulation \cite{Stiles2016b}, complicating the determination of the intrinsic SHE in the NM. Consequently, estimates of the charge-to-spin conversion ratio, namely the spin Hall angle $\theta_{\rm SH}$, and of the spin diffusion length \sdl vary by more than one order of magnitude for the same metal \cite{Sinova2015}. In order to gain fundamental insight into the mechanisms leading to spin accumulation and optimize the spintronic devices that utilize the SHE, it is therefore essential to study the SHE directly in the NM layers.

A straightforward method to detect the SHE is to measure the resulting spin accumulation through the magneto-optical Kerr effect (MOKE). This technique has been employed to reveal the SHE in semiconductors, where \sdl is of the order of a few $\mu$m and the spin accumulation can be laterally resolved by polarization-sensitive MOKE microscopy \cite{Kato2004,Sih2005}. The situation is much more difficult in the case of a metallic conductor such as Pt, where \sdl is just a few nm.
Not only is it unfeasible to detect the lateral spin accumulation with optical wavelengths, also the magnitude of the spin accumulation scales with \sdl and is, despite the relatively large $\theta_{\rm SH}$, one to two orders of magnitude smaller compared to semiconductors.
Nevertheless, first experiments have been performed to directly study the spin accumulation in heavy metal films by MOKE. A report by \citet{Erve2014} claims a spin accumulation signal on an 8~nm thick film of $\beta$-W and, although less clear, on a 20~nm thick film of Pt. The apparent sign change of the observed effect is argued to prove that the polarization rotation, amounting to $3 \cdot 10^{-4}$~rad for $\beta$-W, is due to the SHE.
A follow-up study by \citet{Riego2016}, however, does not support the above conclusions. In that work, magneto-optic ellipsometry measurements with a Kerr rotation detection limit of $10^{-7}$~rad show that any observed current-induced effect is related to a change of the reflectivity of the sample caused by Joule heating. More recently, \citet{Su2017} came to similar conclusions, arguing that MOKE detection would require a current density $j$ larger than $10^8$~\Acm. In fact, all three studies \cite{Erve2014,Riego2016,Su2017} used $j$ in the order of $10^5$~\Acm, which leads to an estimated Kerr rotation of the order of $10^{-9}$~rad \cite{Su2017}, five orders of magnitude smaller than the rotation reported initially \cite{Erve2014}. Alternative optical approaches to detect the spin accumulation in NM include Brillouin light scattering \cite{Fohr2011} and second harmonic generation \cite{Pattabi2015}. Using the latter technique, \citet{Pattabi2015} reported evidence of current induced spin accumulation in Pt, demonstrating also the feasibility of time-resolved studies.
However, the interpretation of the second harmonic signal is not as straightforward as for MOKE.

In this work, we demonstrate the unambiguous detection of the SHE in heavy metals using linear magneto-optical measurements combined with current modulation techniques. We use scanning MOKE microscopy with a sensitivity of $5 \times 10^{-9}$~rad to detect the spin accumulation at the surfaces of Pt and W wires caused by the SHE. Additionally, we perform \ai linear response calculations of the SHE and magneto-optical Kerr rotation \cite{Oppeneer01} caused by the spin accumulation. Comparison of the experimental data with the \ai MOKE calculations provides quantitative values for the spin accumulation, the spin diffusion length, and the spin Hall angle of Pt. These measurements provide a reliable estimate of the SHE and spin diffusion parameters in a NM, without an adjacent FM.

\begin{figure}[t]
	\includegraphics[width=8.0cm]{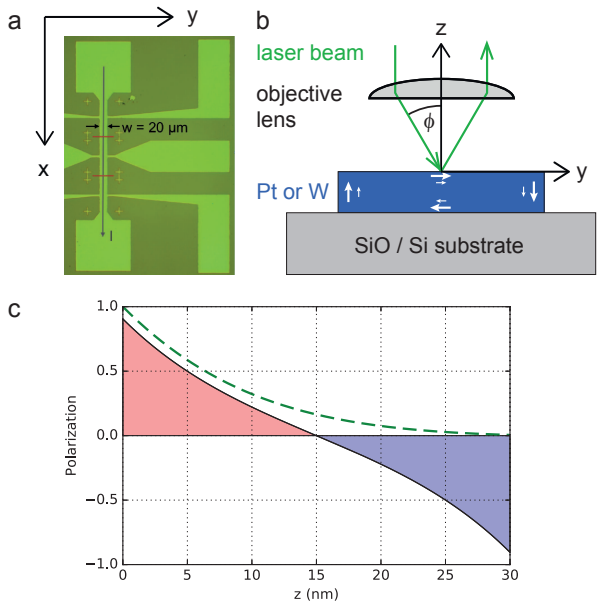} 
	\caption{(a) Microscope image of a Pt Hall bar with the current running top-down (parallel to $x$). (b) Laser beam path through the focusing objective. The arrows in the Pt or W wire represent the spin accumulation. (c) Depth profile of the calculated spin accumulation (shaded) in a 30~nm thick wire for $\sdl=9$~nm. The dashed line represents the depth-dependent sensitivity of MOKE.}
	\label{schematic}
\end{figure}

Our samples are lithographically-patterned Hall bars of Pt and W with line widths of 10 and 20~$\mu$m [Fig.~\ref{schematic}(a)].
The Pt films, with thicknesses ranging between 5 and 60 nm, and 10~nm thick W films are deposited by sputtering on oxidized Si substrates. Four-terminal measurements show that the resistivity of Pt varies between $\rho=27$ and 16~$\mu\Omega$cm with increasing thickness, whereas the resistivity of W is $\rho = 164$~$\mu\Omega$cm, indicative of $\beta$-phase W. For the MOKE measurement, a laser beam with wavelength $\lambda=514$~nm is focused to $\approx 1$~$\mu$m spot size onto the sample, which is mounted on a piezo scanner. A sine-modulated current with variable amplitude up to $j=1.5\cdot10^7 \Acm$ runs through the central conductor, inducing edge spin accumulation [arrows in Fig.~\ref{schematic}(b)]. The resulting light polarization rotation is measured using a sensitive detection scheme comprising a polarization-splitting Wollaston prism and a balanced photodetector. Half of the beam is sent onto a photodiode for measuring changes of the reflected intensity. Both signals are measured by lock-in amplifiers that record the fundamental frequency of the Kerr rotation amplitude and the second harmonic contribution of the reflected intensity. More details about the setup are given in Ref.~\onlinecite{supplementary}.

In a longitudinal MOKE measurement we detect the accumulation of spins along the in-plane $y$ direction transverse to the electric current flowing along $x$, as illustrated in Fig.~\ref{schematic}(a,b). A consequence of the transverse spin current generated by the SHE is the accumulation of spins of opposite sign at opposite interfaces. We therefore detect the superposition of the polarization rotation from spins accumulated at the top and bottom interfaces, drawn as shaded areas in Fig.~\ref{schematic}(c). For quantitative analysis one needs to take into account the light attenuation of the probing laser beam in the conductive material, drawn as dashed line in Fig.~\ref{schematic}(c), as modeled by depth-dependent MOKE calculations \cite{Traeger1992, Hamrle2002}. At the same time, the material-dependent \sdl determines the spatial distribution and the amount of spin accumulation \cite{Zhang2000}, which will be reduced for films of thickness $t$ comparable or smaller than \sdl. These effects will lead to a saturation of the spin accumulation detected by MOKE for films of sufficient thickness. From a thickness-dependent study we can therefore extract the intrinsic spin accumulation and \sdl of a single Pt film, as opposed to the usual Pt/FM bilayer.

\begin{figure}[t]
	\includegraphics[width=\columnwidth]{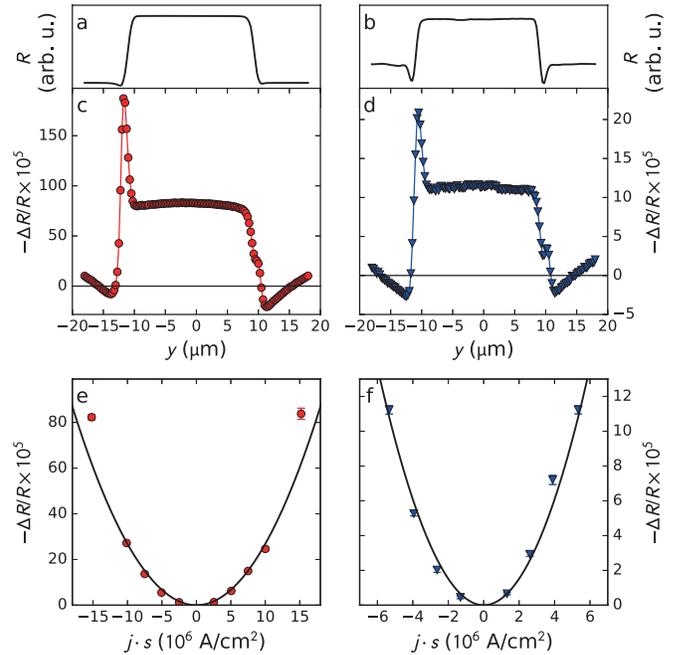}
	\caption{(a,b) Line scans of the optical reflectivity across 20~$\mu$m-wide wires of (a) 15~nm Pt and (b) 10~nm W. (c,d) Thermally induced change of the reflectivity $-\Delta R/R$ for (c) Pt at $j = 1.5 \cdot 10^7$~\Acm and (d) W at $j = 5.3 \cdot 10^6$~\Acm. (e,f) Current dependence of $-\Delta R/R$ for (e) Pt and (f) W. The solid lines are fits to a $j^2$ function. Points left of the origin were measured with the reversed optical path, with $s$ denoting the sign of the incidence angle of the laser beam.}
	\label{linescans}
\end{figure}

We first describe the effect of current injection on the optical reflectivity of Pt and W, which is at the origin of controversial MOKE experiments \cite{Erve2014,Riego2016,Su2017}. Figure~\ref{linescans} displays the results from scanning the laser beam in the $y$-direction across the Hall bar while injecting a sinusoidal current. Both materials, Pt and W, exhibit a change of the reflected intensity $R$, plotted in Fig.~\ref{linescans}.
The relative change $\Delta R/R$, measured in the intensity photodiode as second harmonic of the sine current normalized by the average $R$,
scales as $j^2$ and has the same sign in Pt and W. We therefore assign it to temperature-induced changes of the reflectivity \cite{Favaloro2015} due to Joule heating. Using the relationship $\Delta R / R = c_{TR} \Delta T$, where $c_{TR}=-0.58\cdot 10^{-4}$~K$^{-1}$ for Pt \cite{Favaloro2015}, we estimate a temperature raise $\Delta T$ ranging from 0.2 to 14.3~K as $j$ increases from $2.5 \cdot 10^6$ to $1.5 \cdot 10^7 \Acm$.

\begin{figure}[t]
	\includegraphics[width=\columnwidth]{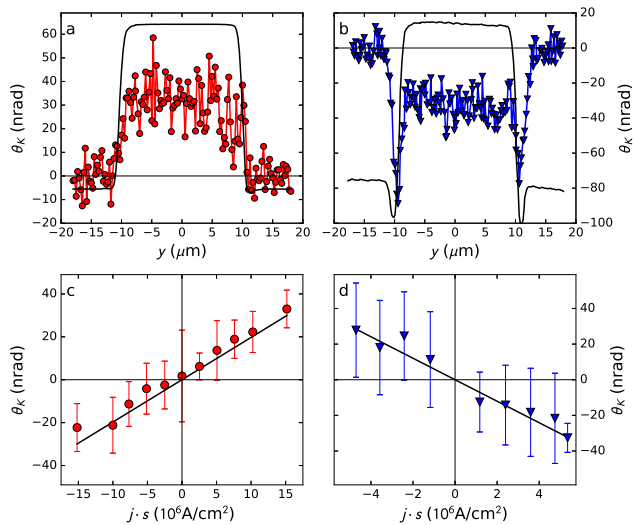}
	\caption{SHE induced Kerr rotation. (a) Line scan of $\theta_{\rm K}$ across a 20~$\mu$m wide, 15~nm thick Pt wire at a current density $j = 1.5 \cdot 10^7$~\Acm and (b) a 10~nm thick W wire at $j=5.3 \cdot 10^6$~\Acm. (c,d) $\theta_{\rm K}$ as a function of $j$. The data points represent line scan averages; statistical error bars from averaging multiple line scans are indicated. The solid lines are linear fits to the data. Points left of the origin were measured with the reversed optical path, $s$.}
	\label{current}
\end{figure}

Crucial for our study is the Kerr rotation, which is measured as the voltage output of the balanced detector at the fundamental frequency of the driving current and calibrated using a half-wave plate. Figure~\ref{current}(a,b) show the Kerr rotation angle $\theta_{\rm K} $ measured on 15~nm thick Pt and 10~nm thick W during sinusoidal current injection.
We observe a clear Kerr rotation signal from the surface of the conducting wires, which is of the order of a few tens of nrad and has opposite sign in Pt and W.
Apart from spurious edge effects, which we attribute to irregular light reflections at the sample boundaries, $\theta_{\rm K} $ is approximately constant over the wire surface, consistent with the spin accumulation picture in Fig.~\ref{schematic}(b). Moreover, we find that $\theta_{\rm K} $ varies linearly with the applied current, as shown in Fig.~\ref{current}(c,d).

Further evidence that $\theta_{\rm K}$ stems from the accumulated spins at both interfaces and is thus a direct consequence of the SHE in the heavy metal layer comes from the following considerations. First, the sine modulation employed here allows for the harmonic separation of different signal contributions, notably the change of the optical reflectivity proportional to $j^2$ (Fig.~\ref{linescans}) and the linear dependence of $\theta_{\rm K}$ on $j$ (Fig.~\ref{current}). In contrast, current switching by square wave modulation, as employed in previous studies \cite{Erve2014,Riego2016,Su2017}, cannot distinguish these effects. We verified that even a slight mismatch between the amplitude of positive and negative current pulses leads to large spurious thermal signals at the fundamental modulation frequency, as discussed also in Ref.~\onlinecite{Su2017}.
Additionally, we implemented an automatic relay scheme that physically inverts the current flowing in the samples and allows us to average out any remaining thermal artifact \cite{supplementary}. Second, control measurements on an Al wire did not result in a detectable Kerr rotation \cite{supplementary}, as expected for a light metal with a minute SHE and large \sdl \cite{Valenzuela2006}. Third, in the longitudinal Kerr geometry chosen here, we are sensitive to spin signatures in the $yz$ scattering plane, \ie, to the in-plane components along $y$ and perpendicular ones along $z$. The two contributions exhibit odd and even symmetry upon inversion of the light incidence angle, for the $y$ and $z$ components, respectively. Figure~\ref{current}(c) and (d) report $\theta_{\rm K} $ measured with opposite angles of incidence in the the two halves of each diagram, which prove that the Kerr rotation changes sign by reversing the optical path of the laser beam. The data are fitted by a line that, within error bars, intersects the origin.
This demonstrates the absence of any thermally induced signal that could be introduced after the switching relay, and excludes the presence of a polar contribution, \ie, a magnetization along $z$.
By rotating the sample by $90^{\circ}$ relative to the laser polarization, we also exclude a magnetization along $x$ \cite{supplementary}. We therefore conclude that our signal results uniquely from the in-plane spin accumulation along $y$.

\begin{figure}[!t]
	\includegraphics[width=\columnwidth]{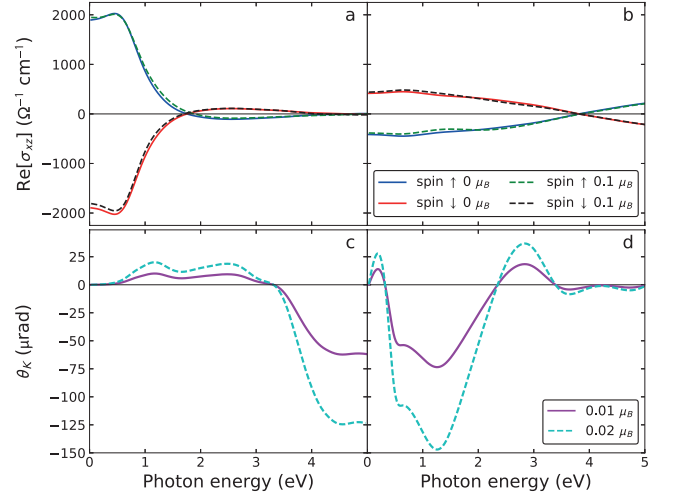}
	\caption{\textit{Ab initio} spin-resolved Hall conductivity Re[$\sigma_{xz}^{y} (\omega )$] as a function of photon energy $\hbar\omega$ of (a) Pt and (b) W. The influence of a spin accumulation on the calculated Re[$\sigma_{xz}^{y} (\omega )$] is shown by the dashed curves (for an induced magnetization of $M^y=0.1$\,$\mu_B$ per atom). (c,d) Calculated longitudinal Kerr rotation spectrum for s-polarized light incident at $37^{\circ}$ in the $yz$ plane, for $M^y=0.01$\,$\mu_B$ (magenta line) and $M^y=0.02$\,$\mu_B$ (dashed cyan line) for Pt and W, respectively.}
	\label{theory}
\end{figure}

To relate the measured Kerr rotation to the amount of accumulated spins we performed \ai calculations of the SHE and of the MOKE due to accumulated spins in Pt and W. We use the density-functional theory framework in the local spin-density approximation to compute the relativistic electronic structure, and employ the linear-response theory to calculate the spin- and frequency dependent Hall conductivity $\sigma_{xz}^{y} (\omega )$ (with $y$ indicating
the spin quantization axis) \cite{oppeneer04} as well as the off-diagonal and diagonal optical conductivities, $\sigma_{ij} (\omega)$ \cite{Oppeneer01}. The DC spin Hall conductivity is given by $\sigma_{xz}^{\rm SH} = {\rm Re}[\sigma_{xz}^{\uparrow} (\omega )- \sigma_{xz}^{\downarrow} (\omega )]/2$, for $\omega \rightarrow 0$. The calculated Re[$\sigma_{xz}^{y} (\omega )$] conductivities of fcc Pt and bcc $\alpha$-W are shown in Fig.~\ref{theory}(a,b); more details are given in Ref.~\onlinecite{supplementary}. The spin-dependent Hall conductivities are antisymmetric in the spin projection and the DC spin Hall conductivities of Pt and W have opposite sign. To investigate the possibility of a feedback effect on the SHE due to the spin accumulation, we calculated the spin-dependent Hall conductivities in the presence of an induced magnetization [dashed curves in Fig.~\ref{theory}(a,b)] and find that its influence is negligible.
The calculated spin Hall conductivity of Pt, $\sigma_{xz}^{\rm SH} ({\rm Pt}) = 1890$ $\Omega^{-1}{\rm cm}^{-1}$ is furthermore in agreement with previous calculations
\cite{guo08,Tanaka2008,Wang2016} and well within the range of measured values ($\sim$$1900 \pm 500$~$\Omega^{-1}{\rm cm}^{-1}$ \cite{Sagasta2016}). Next, we compute the longitudinal Kerr rotation $\theta_{\rm K}$ spectrum for s-polarized light as a function of the induced $y$-magnetization ($M^y$) in Pt and W. The results are shown in Fig.~\ref{theory}(c,d), where, for better visibility, we show the curves corresponding to $M^y= 0.01$ and 0.02~$\mu_B$ per atom, having verified that $\theta_{\rm K}$ scales linearly with $M^y$. This information will be used in the following, where we limit the discussion to Pt, for which there is no ambiguity of crystal structure.

We use the \ai calculated $\sigma_{xz}^{\rm SH}$ and MOKE/$\mu_B$ to compute the spin accumulation in Pt, then compute the theoretical $\theta_{\rm K}$, and compare it with our experiment. By solving the drift-diffusion equation for spins polarized parallel to $y$ and $-y$ for a film of thickness $t$ \cite{Zhang2000}, we obtain the spin accumulation potential
\begin{equation}
\label{eq:spin_pot}
V^y_s(z)= 4 \sdl \sigma_{xz}^{\rm SH} \, \rho^2 j \sinh\left(\frac{t-2z}{2\sdl}\right)\left[\cosh\left(\frac{t}{2\sdl}\right)\right]^{-1}.
\end{equation}
The induced magnetization profile (in $\mu_B$) can then be calculated as
$M^y(z)=\frac{1}{2}eV^y_s(z) D(E_{\rm F})F $,
where $e$ is the electron's charge, $D(E_{\rm F})=1.67$ states/eV is the \ai calculated density of states at the Fermi energy,
and $F=2$ is the Stoner enhancement factor of Pt \cite{Gunnarsson1976}. Using the depth sensitivity of longitudinal MOKE \cite{Traeger1992} we derive the Kerr rotation expected in a measurement of a thin film \cite{supplementary},
\begin{eqnarray}
\label{eq:meask}
\theta_{\rm K} &=& \frac{\sdl \sigma_{xz}^{\rm SH} \, \rho^2 j D(E_{\rm F})F \, \mathrm{e}^{\frac{t}{2\sdl}}}{\cosh(\frac{t}{2\sdl})} \times \nonumber \\
& {\rm Re}& \left\{ \Phi_{\rm K}^{bulk}\kappa \bigg( \frac{(\mathrm{e}^{-\kappa^-t}-1)\mathrm{e}^{-\frac{t}{\sdl}}}{\kappa^-}-\frac{\mathrm{e}^{-\kappa^+t}-1}{\kappa^+}\bigg)
\right\},
\end{eqnarray}
where $\Phi_{\rm K}^{bulk}$ is the bulk complex Kerr effect, and we have defined $\kappa= (4\pi i \bar{n} \cos\psi )/\lambda $, $\cos \psi= (1-\sin^2\phi_i/\bar{n}^2)^{1/2}$, with $\bar{n}$ the complex index of refraction, $\phi_i$ the angle of incidence, and $\kappa^{\pm}=\kappa \pm 1/\sdl$.
From our \ai calculations we obtain values for $\sigma_{xz}^{\rm SH}$, $\bar{n}$, and $\Phi_{\rm K}^{bulk}$  \cite{supplementary}, while other quantities are given from the experiment ($\rho$, $t$, $j$, $\phi_i$).

\begin{figure}[!t]
	\includegraphics[width=8.0cm]{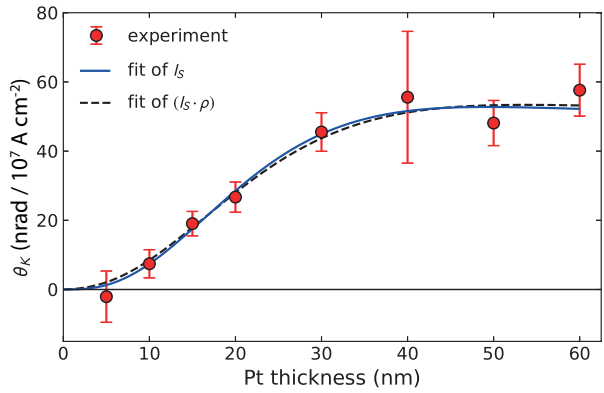}
	\caption{Kerr rotation vs.\ Pt thickness for $j = 10^7$~\Acm. The symbols are obtained from linear fits of $\theta_{\rm K}$ as a function of $j$, as shown in Fig.~\ref{current}. The solid curve is a fit using Eq.\ (\ref{eq:meask}) and $\sigma_{xz}^{\rm SH}= 1880$~$\Omega^{-1}{\rm cm}^{-1}$, $\sdl=11.4$~nm as free parameters. The dashed line is a fit with constant $\sdl\rho = 2.6$~f$\Omega$m$^2$.}
	\label{thickness}
\end{figure}

Figure~\ref{thickness} compares the experimental and computed $\theta_{\rm K}$ of Pt as a function of film thickness for a current density $j=10^7$~\Acm. We observe that, after an initial increase, $\theta_{\rm K}$ saturates for $t\agt30$~nm. This behavior can be understood when one considers two effects, the limited probing depth of our optical measurements and the opposite spin accumulation at the top and bottom interfaces due to the SHE. The solid line represents a fit of $\theta_{\rm K}$ computed using Eq.\ (\ref{eq:meask}) taking the average resistivity $\rho = 20.6$~$\mu\Omega$cm from the experiment and $\sigma_{xz}^{\rm SH}$ and \sdl as free parameters. The fit gives $\sigma_{xz}^{\rm SH}= 1880$~$\Omega^{-1}{\rm cm}^{-1}$, in excellent agreement with theory, and \sdl=\,11.4~nm. These values represent, to our knowledge, the first estimate of the intrinsic \sdl and $\sigma_{xz}^{\rm SH}$ of Pt, independently from the proximity with other metals. The spin Hall angle obtained from this fit is $\theta_{\rm SH}=2\sigma_{xz}^{\rm SH}\rho = 0.08 \pm 0.02$, where the error accounts for the thickness dependence of $\rho$ \cite{supplementary}. Our \sdl is significantly larger than that reported for NM/FM bilayers ($\sdl \approx 1-2$~nm \cite{Sinova2015}), and closer to that obtained by measuring spin absorption in nonlocal devices ($\sdl = 2 - 11$~nm, depending on $\rho$ and temperature \cite{Niimi2013,Sagasta2016}). We note that our estimate assumes constant $\sigma_{xz}^{\rm SH}$, $\rho$, and $\sdl$ parameters, consistently with the derivation of Eq.~(\ref{eq:spin_pot}). However, if Elliott-Yafet spin relaxation dominates in Pt, one expects $\sdl \propto \rho^{-1}$  \cite{Rojas2014,Nguyen2016,Sagasta2016}. If we take this constraint into account together with the experimental variation of $\rho$, our fit gives $\sigma_{xz}^{\rm SH}= 1790$~$\Omega^{-1}{\rm cm}^{-1}$ and $\sdl\rho = 2.6$~f$\Omega$m$^2$ [dashed line in Fig.~(\ref{thickness})]. The latter value is larger than $\sdl\rho = 0.6 - 1.3$~f$\Omega$m$^2$ reported by other techniques \cite{Niimi2013,Rojas2014,Nguyen2016,Sagasta2016}, as discussed in Ref.~\onlinecite{supplementary}.

Finally, by using the proportionality constant between $\theta_{\rm K}$ and the induced magnetic moment \cite{supplementary}, we estimate that the magnetization detected by MOKE at a current density $j = 10^7$~\Acm in the thicker Pt films ($t\geq 40$~nm) corresponds to $(5.0 \pm 0.6)\cdot 10^{-5} \mu_B$/atom in the topmost layer, whereas the average magnetization in the upper half of the films is $(2.0 \pm 0.2)\cdot 10^{-5} \mu_B$/atom.

In conclusion, we have used MOKE microscopy combined with \ai calculations of MOKE and spin Hall conductivity to measure the spin accumulation caused by the SHE in Pt and W thin films. Our results demonstrate the feasibility of characterizing the SHE in NM using magneto-optical methods, independently of the presence of another metal, opening the way to map the spatial and temporal evolution of the spin accumulation and diffusive dynamics in materials with strong spin-orbit coupling and small \sdl.

We acknowledge funding by the Swiss National Science Foundation (grants No.\ 200021-153404, 200020-172775), the Swedish Research Council (VR), the K.\ and A.\ Wallenberg Foundation (grant No.\ 2015.0060), and the Swedish National Infrastructure for Computing (SNIC).

\clearpage
\onecolumngrid{
\center
\bf{
	{\Large Supplemental Material} \\
	\bigskip \bigskip
}
}
	
\setcounter{equation}{0}
\setcounter{figure}{0}
\setcounter{table}{0}
\setcounter{page}{1}
\makeatletter
\renewcommand{\thesection}{SM \arabic{section}}
\renewcommand{\theequation}{S\arabic{equation}}
\renewcommand{\thefigure}{S\arabic{figure}}
\renewcommand{\bibnumfmt}[1]{[S#1]}
\renewcommand{\citenumfont}[1]{S#1}
\renewcommand{\thetable}{S\arabic{table}}


\section{Experimental setup}
\label{sect:exp:setup}
\begin{figure}[bh]
	\includegraphics[width=11.5cm]{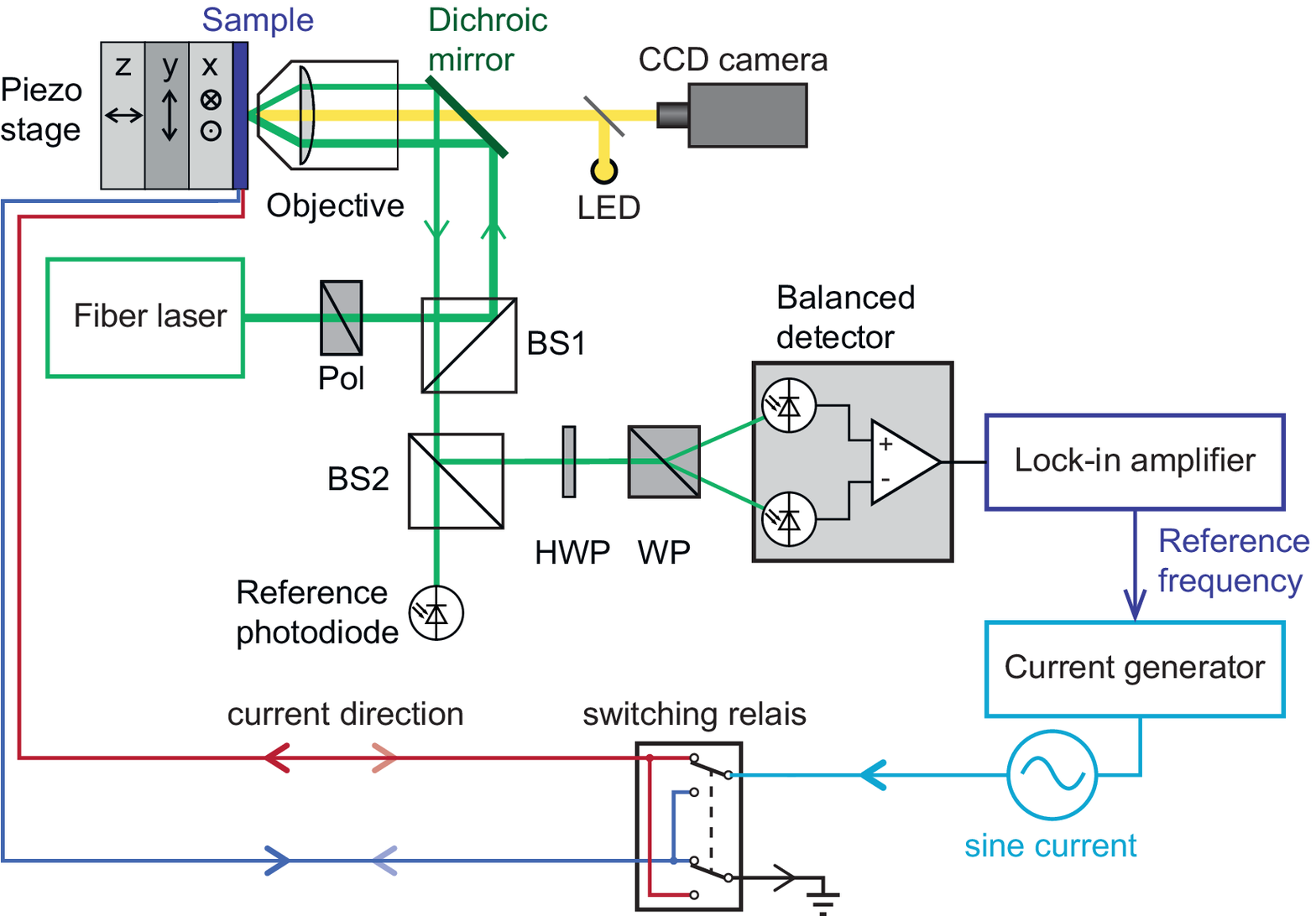}
	\caption{Schematic drawing of the scanning Kerr effect microscope set-up. The laser beam passes the polarizer (Pol) and first beamsplitter (BS1), is reflected by a dichroic mirror, and enters the objective to be focused onto the sample. The reflected beam is collimated by the same objective, split into two beams at the second beamsplitter (BS2), and its intensity is monitored in the reference photodiode. Light polarization rotation is measured in a second detection branch, comprising of a half wave plate (HWP), Wollaston prism (WP), and balanced photodiode detector.}
	\label{setup}
\end{figure}

Figure~\ref{setup} shows a detailed schematics of our experimental setup.
The sample is mounted on a piezo-controlled $xyz$--stage for scanning in the $xy$~plane in the laser focus, $z$ being the sample's surface normal. For measuring the longitudinal magneto-optic Kerr effect (MOKE), we use a fiber laser (Origami 10-05 from Onefive GmbH) at $\lambda=514$~nm wavelength (2.41~eV photon energy), attenuated to a power of $\approx 120~\mu$W for Pt and $\approx 20~\mu$W for W. The beam is incident in the $yz$ plane, which makes the longitudinal MOKE measurement sensitive to magnetic moments along $y$. The angle of incidence of $\phi_i \approx 37^\circ$ is achieved by a parallel displacement of the laser beam from the central axis upon entering the objective. As we use s-polarized light, we do not get contributions from the transverse MOKE. The beam is focused on the sample to a spot size of about 1~$\mu$m by the $100\times$ microscope objective with numerical aperture $\mathrm{NA}=0.9$.

The intensity and polarization state of the reflected laser beam are measured by a reference photodiode and a balanced detection setup, respectively. The latter comprises a Wollaston prism which splits the light into two beams polarized linearly perpendicular to each other, and a balanced detector to measure their intensity difference.
This difference, which is proportional to the polarization rotation, is initially adjusted to zero by rotating a half-wave plate in front of the Wollaston prism. Using a calibration procedure in which we deliberately rotate the polarization axis with the half-wave plate, we determine the proportionality constant between detector output voltage and polarization rotation angle. The signal from the balanced detector is fed into a lock-in amplifier, yielding the spin Hall effect (SHE)-induced Kerr rotation by demodulating at the first harmonic.
As the expected rotation signals are very small, we average several line scans of typically 15 minutes integration time each over a period of 24-48 hours, for each data point shown in Fig.~3a,b of the main text.
For measuring changes in the sample's reflectivity, we monitor the reference photodiode's AC component at the second harmonic of the modulation frequency.

The sinusoidally modulated current of frequency 2030~Hz is generated in a voltage-controlled current source driven by the lock-in reference output and passed through the central conductor of the Hall bar patterned on the sample, parallel to the $x$ direction. The peak amplitude of the current density is indicated as $j$ throughout the manuscript. A relay switch between the current source and the sample reverses the direction of current flow between individual line scans. This procedure allows us to detect the presence of possible asymmetries in the sine-modulated current and, by means of subtracting signals measured with opposite relay switch settings, exclude these effects from the analysis of the Kerr rotation. We found that these asymmetries are negligible when using sine current modulation and lock-in detection, but may become predominant for rectangular pulses of slightly different amplitudes for positive and negative current.

\section{Sample fabrication and electrical characterization}
\label{sect:exp:samples}
Pt and W thin films were deposited on oxidized Si(100) wafers by magnetron sputtering. The substrates were cleaned in a 50~$^{\circ}$C heated ultrasonic bath first in acetone followed by isopropanol, for 10~minutes each. Sputter deposition with base pressure of $2 \cdot 10^{-8}$~Torr was performed at room temperature. Prior to deposition, further cleaning of the substrate surface was done by argon RF sputtering during 60~s at an Ar partial pressure of 10~mTorr. The Pt samples were deposited in a 0.3~mTorr Ar atmosphere at a growth rate of 1.57~\AA/s while rotating the sample holder at 30~rpm. The W films were deposited at a growth rate of 0.041~\AA/s in a 5.0~mTorr Ar atmosphere while being rotated at 30~rpm. Finally, the Pt and W thin films were patterned into Hall bar structures (Fig.~\ref{Hallbar}) by photolithography and Ar ion milling.

\begin{figure}[htb]
	\includegraphics[width=6cm]{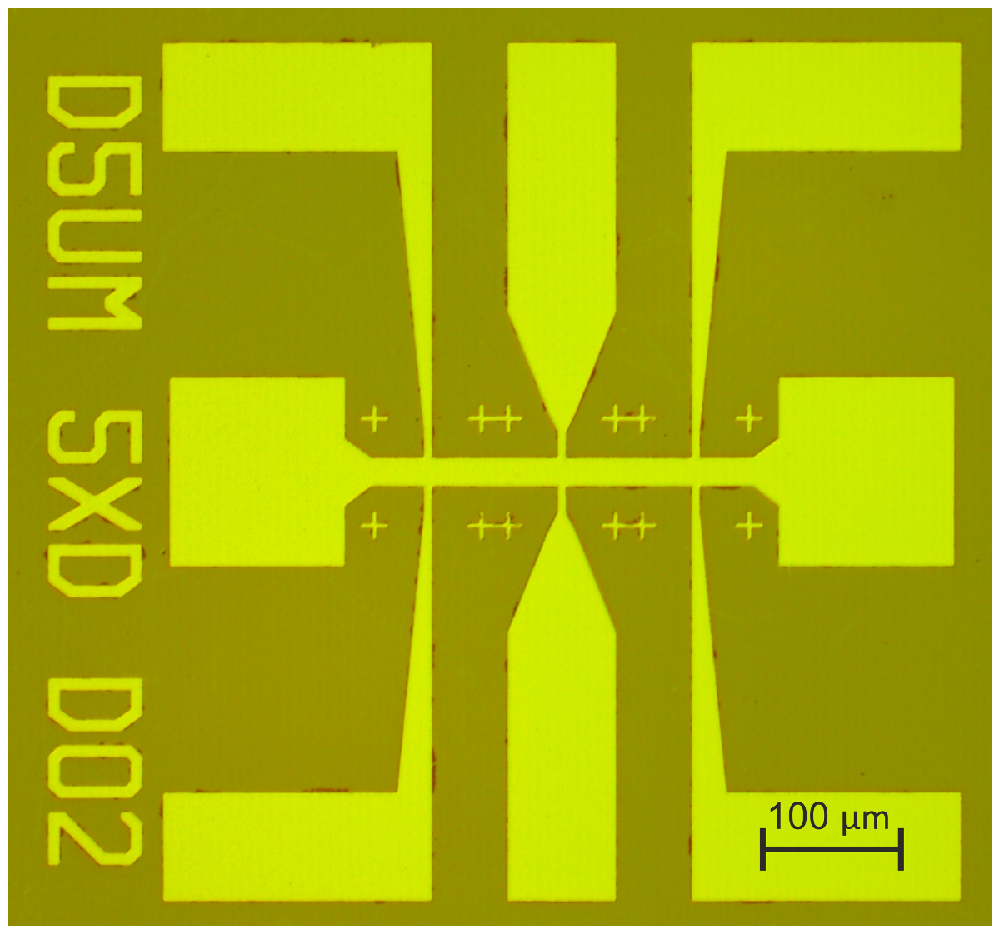}
	\caption{Image of a photolithographically patterned 10 nm thick Pt sample. Current flows through the horizontal bar, the additional contacts are used to measure the resistance.}
	\label{Hallbar}
\end{figure}

\begin{figure}[h]
	\begin{minipage}[c]{0.55\textwidth}
		\begin{tabular}{@{}lrrr@{}} \toprule[0.05em]
			Sample  & \hspace{3mm} R ($\Omega$) & \hspace{3mm}$\rho$ ($\mu \Omega \mathrm{cm}$)  & \hspace{3mm}$\rho_{high}$ ($\mu \Omega \mathrm{cm}$) \\
			\midrule[0.025em] 
			Pt (5 nm) & 519.6 & 27.0 & 27.2 \\
			Pt (10 nm) & 208.1 & 21.8 & 22.1 \\
			Pt (15 nm) & 137.5 & 20.8 & 21.5 \\
			Pt (20 nm) & 89.0 & 19.9 & 20.6 \\
			Pt (30 nm) & 60.3 & 18.9 & 19.7 \\
			Pt (40 nm) & 37.9 & 15.2 & 16.3 \\
			Pt (50 nm) & 34.3 & 16.8 & 18.4 \\
			Pt (60 nm) & 28.6 & 16.9 & 18.9 \\
			W (10 nm) & 1690.6 & 164.0 & 164.3 \\
			heated W (10 nm) & 1203.2 & 115.5 & 116.7 \\
			Al (25 nm) & 22.4 & 5.9 & 6.0 \\
			\bottomrule[0.05em]		
		\end{tabular}
		\setcounter{figure}{0}
		\renewcommand\figurename{TABLE}
		\caption{List of measured samples, their resistance $R$ and resistivity $\rho$ measured at a current density of $0.5 \cdot 10^{6} $ A/cm$^{2}$ as well as the resistivity $\rho_{high}$ measured at $5.0 \cdot 10^{6} $ A/cm$^{2}$. Four-point probe measurement were done on Hall bar structures that are 200~$\mu \mathrm{m}$ long.}	
		\label{resistance}	
	\end{minipage}
	\hspace{0.5cm}
	\begin{minipage}[c]{0.35\textwidth}
		\centering
		\includegraphics[width=4.8cm]{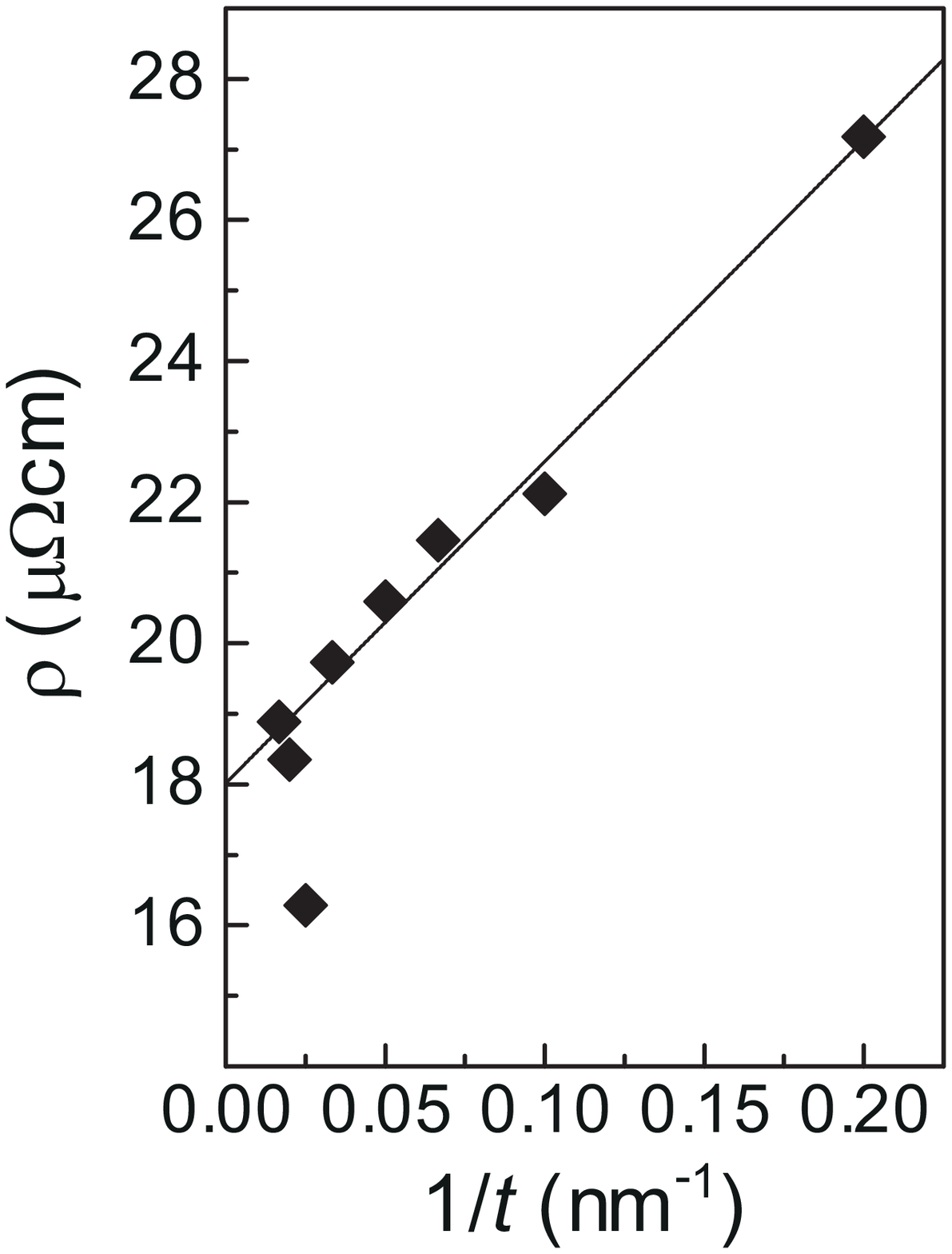}
		\setcounter{figure}{2}
		\vspace{-0.35cm}
		\caption{Resistivity as a function of inverse Pt thickness measured at a DC current density of $5.0 \cdot 10^{6}$ A/cm$^{2}$. The solid line is a linear interpolation of the data, which gives $\rho = (20.6 + 45/t\mathrm{[nm]})$~$\mu\Omega$cm.}
		\label{FigS2rho}
	\end{minipage}
\end{figure}

Four-point probe resistivity measurements were performed on all samples, as listed in Table~\ref{resistance}, using DC current densities ranging from $0.5 \cdot 10^{6}$ A/cm$^{2}$ to $5 \cdot 10^{6}$ A/cm$^{2}$. The resistivity of Pt versus thickness $t$ has an approximate $1/t$ behavior, shown in Fig.~\ref{FigS2rho}, as expected for thin metal films. Also, owing to Joule heating, the resistivity increases slightly with increasing current density. This effect is more noticeable in the thicker films, as the dissipated power is proportional to the sample volume. We use the resistivity measured at high current to model the Kerr rotation of Pt using Eq.~(2) in the main text, noting that a DC current density of $5 \cdot 10^{6}$ A/cm$^{2}$ dissipates the same average power as an AC current density of amplitude $1 \cdot 10^{7}$ A/cm$^{2}$.

Literature values for the resistivity of W can reach $260~\mu \Omega \mathrm{cm}$ for 5.2~nm thick $\beta$-W \cite{Pai2012}, or $200~\mu \Omega \mathrm{cm}$ for 12~nm $\beta$-W, and $110~\mu \Omega \mathrm{cm}$ for annealed 12~nm $\beta$-W \cite{Hao2015}. For $\alpha$-W, values between $21~\mu \Omega \mathrm{cm}$ \cite{Pai2012} and $35~\mu \Omega \mathrm{cm}$ \cite{Hao2015} are found. We therefore conclude that our sample is predominantly $\beta$-W, and that the application of a sine current of 25~mA to a second W sample (denoted as ``heated W'' in Table \ref{resistance}) seems to have heated the sample enough to introduce at least a partial phase change from $\beta$-W to $\alpha$-W.

\section{Orientation of the SHE--induced spin accumulation in platinum}
\begin{figure}[h]
	\includegraphics[width=15cm]{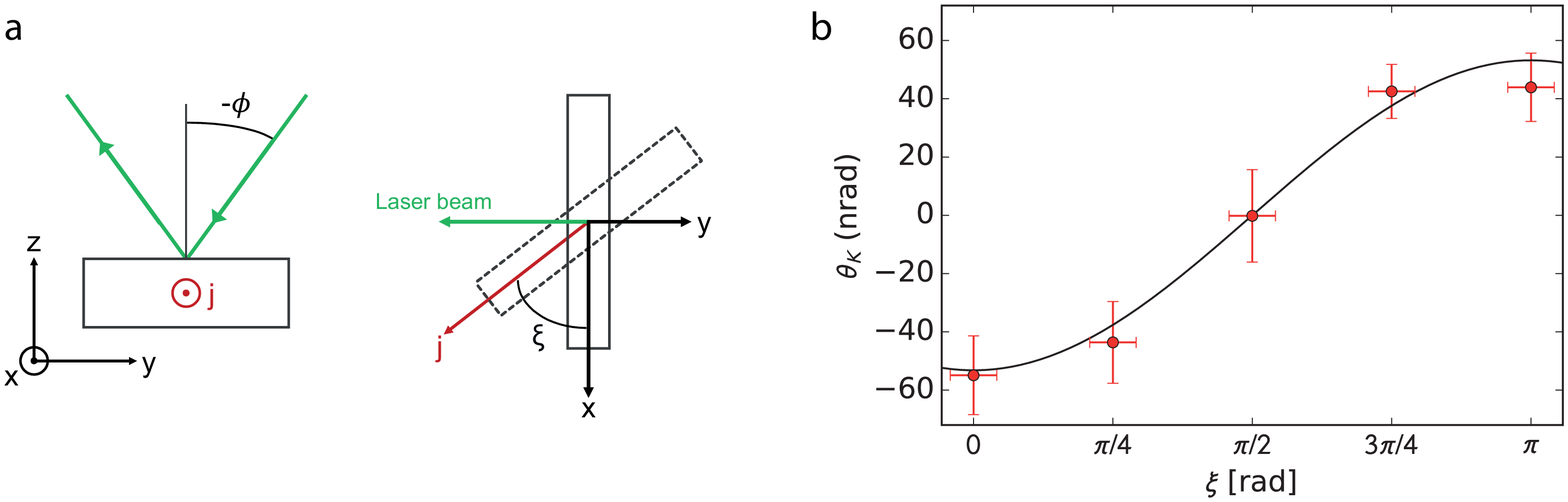}
	\caption{(a) Schematic of the experimental geometry used to measure the angular dependence of the Kerr rotation. (b) $\theta_{\rm K}$ measured as a function of $\xi$ on a 50~nm thick Pt sample with current amplitude $j = 10^{7}$ A/cm$^{2}$. Note that the optical path is reversed with respect to Fig.~1 of the main text.}
	\label{angle_dep}
\end{figure}

In addition to reversing the optical path, we probed the direction of the spin accumulation by rotating the sample with respect to the plane of incidence of the s-polarized laser beam, as shown in Fig.~\ref{angle_dep} (a). The measurements were done on Pt(50 nm) at a current density $j = 10^{7}$ A/cm$^{2}$. The dependence of the Kerr rotation $\theta_{\rm K}$ on the angle $\xi$ between the current direction and the $x$ axis is shown in Fig.~\ref{angle_dep} (b). Note that for these measurements we used the reversed optical path geometry (laser beam parallel to $-y$), so that $\theta_{\rm K}<0$ at $\xi=0$. We observe that the Kerr signal increases as $-\cos \xi$ upon rotation of the current direction away from the $x$ axis, as expected for a current-induced magnetization perpendicular to $j$. At $\xi = \pi/2$, the Kerr rotation vanishes, indicating that the magnetization component parallel to the current direction is zero within our detection limit. We further note that the antisymmetry of $\theta_{\rm K}$ upon reversal of the optical path (see Fig.~3 of the main text) indicates that any polar contribution to the Kerr rotation is negligible, consistently with the absence of current-induced magnetization along $z$. Overall, our measurements show that the current-induced magnetization in Pt is directed uniquely in-plane, perpendicular to $j$.

\section{Absence of SHE--induced Kerr rotation in aluminium}

\begin{figure}[ht]
	\includegraphics[width=12cm]{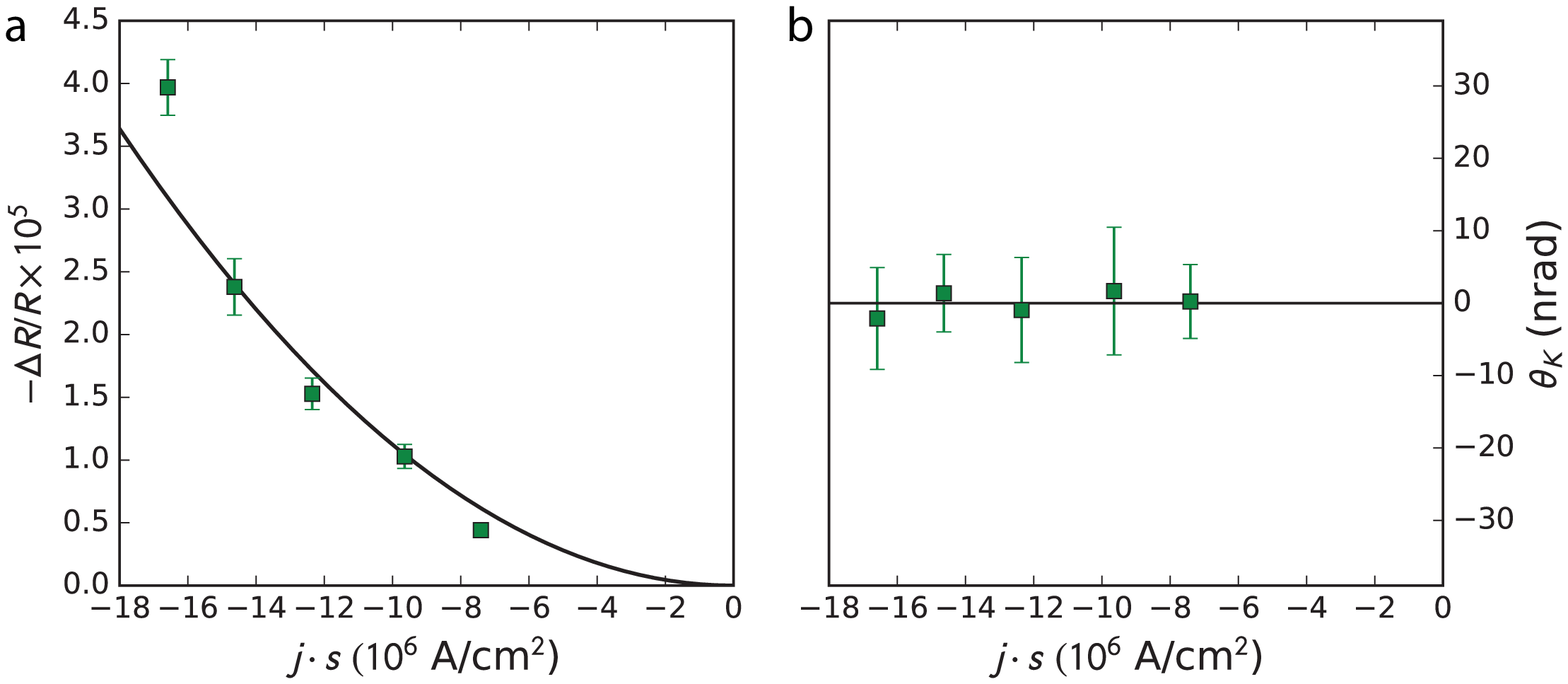}
	\caption{Comparison measurement on Al showing (a) temperature-induced change of the reflectivity and (b) absence of the SHE-induced Kerr rotation as function of the current density $j$. The $x$-scale is multiplied by the sign $s$ of the light incidence angle, for consistency with the data reported in the main text.}
	\label{Al}
\end{figure}

In order to test our measurement procedure, we performed additional measurements on a current stripe of Al, a material which is known to have a very small spin Hall angle $\theta_{\rm SH} \approx 10^{-4}$ and large spin diffusion length $\sdl > 400$~nm \cite{valenzuela2006}. The results are plotted in Fig.~\ref{Al}, where no SHE-induced Kerr rotation could be detected within the 5~nrad detection limit, while the reflectivity change is clearly present. Due to the low resistivity of Al, 6.0 $\mu \Omega$~cm, and the resulting lower Joule heating, the thermal-induced change of the optical reflectivity is not as pronounced as for Pt and W.

\section{Theoretical background}

We provide here the theoretical background of the detection of the SHE in Pt and W thin metal films using MOKE.
Specifically, we employed the linear-response theory combined with the Density Functional Theory (DFT) framework to compute \textit{ab initio} the SHE, the optical response as well as the MOKE response. By comparing the calculated Kerr rotation as a function of film thickness with the experimental data, we obtain values for the spin accumulation, spin Hall angle, and spin diffusion length in Pt. The experiment that we are addressing is schematically depicted in Fig.~1a in the main text. A given current is passed through a thin film of Pt or W and the accumulated magnetization that occurs at the top and bottom surfaces due to the SHE is measured using longitudinal MOKE spectroscopy with the light scattering plane parallel to the $yz$ plane.

To theoretically model the  magneto-optical detection of the SHE described in the main text, we have performed the following calculations:
\begin{itemize}
	\item \textit{Ab initio} calculation of the spin Hall conductivity and conductivity tensor of  Pt and W, giving the spin Hall angle.
	\item Calculation of the spin accumulation at the top and bottom surfaces of the thin films, which is used as input for the \textit{ab initio} calculated spin-resolved conductivity tensor.
	\item \textit{Ab initio} calculation of the L-MOKE signal for uniformly magnetized materials (Pt and W).
	\item Calculation of the depth sensitivity of L-MOKE to obtain the theoretical thickness dependent Kerr rotation for a thin film which can be compared with measured values.
\end{itemize}
We perform the calculations for nonmagnetic fcc Pt and bcc W, with lattice constants $a(\textrm{Pt})=3.9242$\,{\AA} and $a (\textrm{W})=3.1652$\,{\AA}, respectively. The theoretical framework and results of the \textit{ab initio} calculations are detailed in the following.

\section{Calculations of the spin-resolved conductivity response}
All our \textit{ab initio} calculations are performed on the basis of the DFT as implemented in the Augmented Spherical Wave (ASW) electronic structure code \cite{williams79,eyert07}. Since we are interested in magnetic features we use the local spin density approximation (LSDA) with the von Barth-Hedin (vBH) parametrization \cite{von72}. This provides us with the relativistic band structure $\epsilon_{n\bm{k}}$ of any specific material and with the relativistic eigenstates $\vert n \bm{k}\rangle$ of the Kohn-Sham Hamiltonian. We then use linear-response theory to calculate most of the relevant quantities that we need to model the experiment which are all related with the conductivity tensor.
The optical conductivity tensor is calculated using the linear-response expression \cite{oppeneer92,mondal15moke}:
\begin{eqnarray}
\label{eq:conductivity}
\sigma_{\alpha\beta}(\omega) &=& -\frac{i \hbar}{V}\sum_{\bm{k}}\sum_{n\neq n'} \frac{f(\epsilon_{n\bm{k}})-f(\epsilon_{n' \bm{k}})} {\epsilon_{n\bm{k}}-\epsilon_{n^\prime \bm{k}}}
\frac{\langle n^\prime \bm{k}\vert\hat{j}^{\alpha}\vert n\bm{k}\rangle\langle n \bm{k}\vert\hat{j}^{\beta}\vert n^\prime \bm{k}\rangle}{\hbar \omega-\epsilon_{n \bm{k}}+\epsilon_{n^\prime \bm{k}}+{i \hbar }/{\tau}}+\sigma_{_{\rm D}}(\omega)\delta_{\alpha\beta} .
\end{eqnarray}
Here $f(\epsilon_{n \bm{k}})$ is the Fermi-Dirac distribution,
$\hat{j}^{\alpha},\, \hat{j}^{\beta}$ are the current density operators in $\alpha,\, \beta$ direction and $\tau$ is the electron  lifetime.
The term $\sigma_{_{\rm D}}(\omega)$ is the Drude contribution to the conductivity,
\begin{equation}
\label{eq:drude}
\sigma_{_{\rm D}}(\omega)=\frac{\sigma_{_{\rm D}}^0}{1-i\omega\tau_{_{\rm D}}} ,
\end{equation}
where $\sigma_{_{\rm D}}^0$ is the Drude peak amplitude and $\tau_{_{\rm D}}$ is  the Drude relaxation time.

The DC spin Hall conductivity can be obtained from computing the spin-resolved, off-diagonal conductivity as given by Eq.\ (\ref{eq:conductivity}) and taking the limit $\omega \rightarrow 0$ (see Ref.\ \cite{oppeneer04sm}).
Adopting here the geometry that the current is along the $x$ direction and the spin quantization axis along $y$,
the corresponding spin-projected off-diagonal conductivity can be written as \cite{oppeneer04sm,guo08sm}:
\begin{equation}
\label{eq:SHE}
\sigma_{xz}^y=\frac{\hbar}{V}
\sum_{\bm{k},n} f(\epsilon_{n \bm{k}})\Omega_n^y(\bm{k}),
\end{equation}
where we define the Berry curvature (modified by the inclusion of the lifetime) as
\begin{equation}
\label{eq:berry_curv}
\Omega_n^y(\bm{k})=\sum_{n^{\prime} \neq n} \frac{\textrm{Im}[\langle n \bm{k}\vert j_x^y\vert n^\prime \bm{k} \rangle\langle n^\prime \bm{k} \vert j_z\vert n \bm{k} \rangle]}{[\epsilon_{n\bm{k}}-\epsilon_{n^\prime \bm{k}}] [\epsilon_{n\bm{k}}-\epsilon_{n^\prime \bm{k}} +i\hbar/\tau ]} .
\end{equation}
This equation differs from the equation given in Ref.\ \cite{guo08sm} due to the inclusion of the lifetime. The quantity $j_x^y=\frac{1}{2}\{ \sigma_y ,j_x\}$ is the spin current operator for propagation along $x$, $\sigma_y$ is the Pauli spin matrix for spin axis along $y$, and $\{ \cdot , \cdot \}$ stands for the anti-commutator. Note that we adopt the typical coordinate system of experiments involving thin films, where $z$ is the direction perpendicular to the film plane, rather than that used in the theoretical literature of the SHE, where the the off-diagonal conductivity is indicated as $\sigma_{x'y'}^{z'}$, with $z'$ being the spin quantization axis and $y'$ the in-plane direction normal to the current.
The frequency-dependent spin projected conductivity is given by Eqs.\ (\ref{eq:SHE}) and (\ref{eq:berry_curv}) as well, but with $i\hbar/ \tau$ replaced by $\hbar \omega + i\hbar / \tau$.

The numerical implementation of the matrix elements of the full current operators is described in Ref.\ \cite{oppeneer92} and that of the spin-projected current operators in Ref.\ \cite{oppeneer04sm}. The $k$-space integration over the Brillouin zone is performed using the tetrahedron integration method \cite{oppeneer92}\footnote{To perform the $k$-space integration we integrate in the irreducible relativistic wedge of the Brillouin zone over 414720 $k$-points either for Pt and W}. We used an electron relaxation time $\hbar \tau^{-1}=0.02$~Ry (or $\tau=2.42 \times10^{-15}$~s) in all our calculations; this value provides a good description of the off-diagonal conductivity of metals \cite{oppeneer01}.

\begin{figure}[ht]
	\includegraphics[width=9cm]{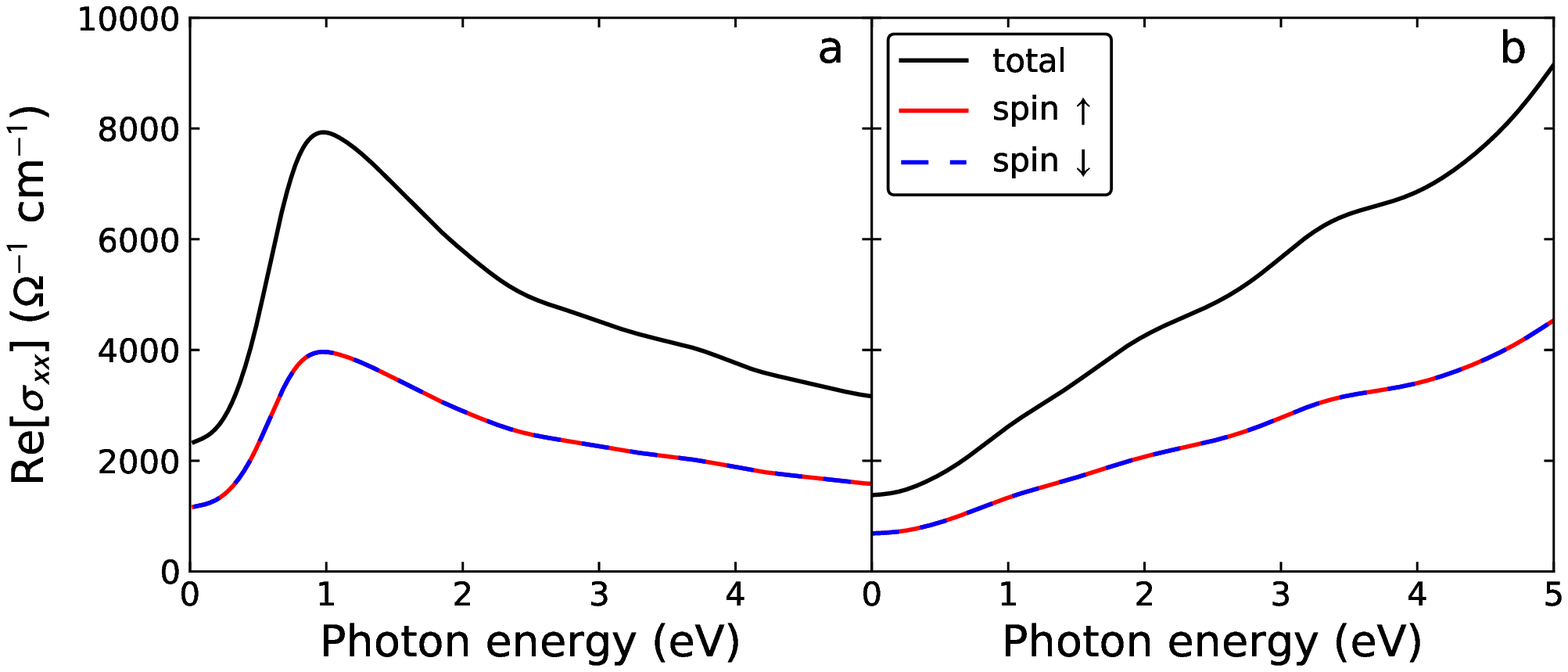}
	\caption{Spin-resolved diagonal optical conductivity calculated \textit{ab initio} for Pt (a) and W (b) as a function of photon energy $\hbar \omega$.}
	\label{fig:diag}
\end{figure}

The \textit{ab initio} calculated results for the real parts of the spin-resolved off-diagonal conductivities of Pt and W (as a function of the photon energy $\hbar\omega$) are shown in Fig.~4 of the main text.
The DC spin Hall conductivity is obtained as the limit for $\omega\rightarrow0$ of the off-diagonal term. Note that the spin-projected off-diagonal conductivity has opposite sign for the spin-up and spin-down contributions, with the total off-diagonal conductivity being zero, as expected for nonmagnetic metals. The spin-resolved diagonal conductivities of Pt and W are shown in Fig.\ \ref{fig:diag}. Note that in contrast to the off-diagonal components,  the contributions of the spin-up and spin-down channels to the diagonal conductivity are equal and add up to a total, nonzero conductivity.
For Pt we obtain results for the spin Hall conductivity in very good agreement with previous calculations \cite{guo08sm,Tanaka2008sm,zhang15,Wang2016sm}. Note that different authors adopt different definitions of the spin Hall conductivity; here, as in Refs.\ \cite{guo08sm,Tanaka2008sm,Wang2016sm}, it is defined as the single spin component conductivity $\sigma_{xz}^{\uparrow}$, or $\sigma_{xz}^{\downarrow}$, i.e., $\sigma_{xz}^{\rm SH} = {\rm Re} [\sigma_{xz}^{\uparrow} -\sigma_{xz}^{\downarrow} ]/2$, whereas in Ref.\ \cite{zhang15} it is represented as $\sigma_{xz}^\uparrow-\sigma_{xz}^\downarrow$, which, given the fact that we investigate nonmagnetic materials ({i.e.}, with zero off-diagonal total conductivity) is equal to $2 \sigma_{xz}^{{\rm SH}}$.
In Table \ref{tab0} we compare the results for $\sigma_{xz}^{\rm SH}$ of Pt as given by the \textit{ab initio} calculations, and compare to the experimental values \cite{zhang15,Sagasta2016sm,Nguyen2016sm}, too. Our computed value is somewhat larger than that obtained by Wang \textit{et al.}\ \cite{Wang2016sm} and smaller but in good agreement with that obtained by Guo \textit{et al.}\ \cite{guo08sm} and Zhang \textit{et al.}\ \cite{zhang15}.

\vspace{0.5 cm}
\begin{table}[ht!]
	\setlength{\tabcolsep}{12pt}
	\begin{tabular}{@{}l l c l@{}}
		\toprule[0.05em]
		Reference & Reported  $\sigma^{\rm SH}\,[\Omega {\rm cm}]^{-1}$ & $\sigma_{xz}^{\rm SH}\,[\Omega {\rm cm}]^{-1}$ & Type \\
		\midrule[0.025em]
		This work & 1890 $\left[ \frac{\hbar}{e} \right]$ & 1890 & theory \\
		This work & $1880\pm 150$ $\left[ \frac{\hbar}{e} \right]$ & 1880 & experiment \\
		Guo {\em et al.}\,\cite{guo08sm} & 2200 $\left[\frac{\hbar}{e}\right]$ &  2200 & theory \\
		Tanaka {\em et al.}\,\cite{guo08sm} & 1000-2000 $\left[\frac{\hbar}{e}\right]$ &  1000-2000 & theory \\
		Zhang {\em et al.}\,\cite{zhang15} & 4370 $\left[\frac{\hbar}{2e}\right]$ & 2185 & theory \\
		Wang {\em et al.}\,\cite{Wang2016sm} & 3200 $\left[\frac{\hbar}{2e}\right]$ & 1600 & theory \\
		Zhang {\em et al.}\,\cite{zhang15} & $1900 \pm 200$ $\left[\frac{\hbar}{2e}\right]$ & 950 & experiment \\
		Nguyen {\em et al.}\,\cite{Nguyen2016sm} & $5900 \pm 200$ $\left[\frac{\hbar}{2e}\right]$ & 2950 & experiment \\
		Sagasta {\em et al.} \cite{Sagasta2016sm} & 1500-3000 $\left[\frac{\hbar}{e}\right]$ & 1500-3000 & experiment\\
		\bottomrule[0.05em]
	\end{tabular}
	\setcounter{table}{1}
	\caption{Comparison of different values of the \textit{ab initio} calculated and measured DC spin Hall conductivity of Pt. The second column reports the values as given in each paper, following different definitions of the spin Hall conductivity. Here we define $\sigma_{xz}^{\rm SH} = (\sigma_{xz}^{\uparrow}-\sigma_{xz}^{\downarrow})/2$, according to which the spin Hall angle, namely the ratio between the spin and charge currents, is given by $\theta_{\rm SH}=(\sigma_{xz}^{\uparrow}-\sigma_{xz}^{\downarrow})/\sigma_{xx}= 2\sigma_{xz}^{\rm SH}/\sigma_{xx}$. The third column reports the spin Hall conductivity according to this definition, which corresponds to the so-called $\left[\frac{\hbar}{e}\right]$ notation.}
	\label{tab0}
\end{table}

\section{Spin accumulation calculations}
We can now use the values calculated for the spin Hall conductivity to compute the spin accumulation at the top and bottom surfaces of the metal films. For this purpose we solve the drift-diffusion equation in the heavy metal layer, as shown in Ref.\ \cite{Zhang2000sm} within a Boltzmann transport equation framework. The spin-dependent potential at the surfaces, due to the SHE, is
\begin{equation}
\label{eq:spin_pot_sm}
V^y_s(z)= 2\sdl \theta_{\rm SH} \rho j_x  \frac{\sinh\left(\frac{t-2z}{2\sdl}\right)}{\cosh\left(\frac{t}{2\sdl}\right)},
\end{equation}
where \sdl is the spin diffusion length in the material, $\theta_{\rm SH}={2\sigma_{xz}^{\rm SH}}/ {\sigma_{xx}}$ the spin Hall angle, $\rho$ is the resistivity, $t$ the film thickness, and $j_x$ the current injected in the material.
The equation is valid when the diffusion length is reasonably smaller than the thickness, which is fulfilled for most of our samples.
Starting from the spin-resolved potential, which can be seen as an effective magnetic field acting on the conduction electrons, we calculate the spin accumulation (in $\mu_B$) making use of the equation
\begin{equation}
\label{eq:mag}
M^{y}(z)=\frac{1}{2}e V_s^y(z)D(E_{\rm F})F,
\end{equation}
where $D(E_{\rm F})$ is the density of states of the material at the Fermi energy and $F$ is the Stoner enhancement factor. As Stoner enhancement factor for Pt we use the value $F=2$ \cite{Gunnarsson1976sm}.
The density of states were calculated using the ASW method within the DFT framework as $D_{\rm Pt}(E_{\rm F})=1.67$ states/eV and $D_{\rm W}(E_{\rm F})=0.34$ states/eV. It is important to mention here that an accurate estimation of the DC longitudinal conductivity is required (which enters in Eq.\ (\ref{eq:spin_pot_sm}) in the spin Hall angle and in $\rho$ which is its inverse).
Even though it would be possible to estimate the bulk conductivity using \textit{ab initio} methods, we prefer to use the experimental DC conductivity ($\sigma_{xx} = 1/\rho$, see Table~\ref{resistance}) as input for our calculations. The reason is that there is often a variation of this quantity encountered for different samples (due to preparation conditions, microstructure, sample purity, and thickness), which is better accounted for by taking the conductivity measured in the experiments.

\section{Calculation of the longitudinal MOKE }
The last quantity that we need to model for the experiment is the prediction of the Kerr rotation for a longitudinal MOKE measurement.
For the calculations of the complex bulk L-MOKE Kerr effect, $\Phi_{\rm K}^{L,bulk} = \theta_{\rm K}^L + i \varepsilon_{\rm K}^L$,
with  $\theta_{\rm K}^L$ the rotation angle and $\varepsilon_{\rm K}^L$ the Kerr ellipticity, we use the following equations, derived in Refs.\ \cite{you96,oppeneer01},
\begin{eqnarray}
\label{eq:kerr}
\! \! \! \! \theta_{\rm K}^L+i\varepsilon_{\rm K}^L &=&\frac{iQ\bar{n}n_0}{\bar{n}^2-n_0^2}\frac{\cos\phi_i\tan\phi_t}{\cos(\phi_i-\phi_t)} \quad \text{(s-polarization)}, \nonumber \\
\! \! \! \! \theta_{\rm K}^L+i\varepsilon_{\rm K}^L &=&\frac{iQ\bar{n}n_0}{\bar{n}^2-n_0^2}\frac{\cos\phi_i\tan\phi_t}{\cos(\phi_i+\phi_t)} \quad \text{(p-polarization)}.
\end{eqnarray}
Here $\phi_i$ is the angle of incidence angle of the beam and $\phi_t$ the angle of transmission in the sample,
$Q$ is the Voigt parameter, defined as
$ Q =i {\epsilon_{xy}}/{\epsilon_{xx}}$, with $\epsilon_{ij}$ elements of the dielectric tensor,
and $\bar{n} = (n^+ + n^-)/ 2$, where $n^{\pm}$ are the refractive indices for $\pm$ circularly polarized eigenmodes in the material.
The dielectric tensor is related to the conductivity tensor as (in
cgs or Gaussian units),
\begin{equation}
{\bm \epsilon}(\omega)=\mathbb{1}+\frac{4\pi i}{\omega}{\bm \sigma}(\omega).
\end{equation}
Using the linear response expression (\ref{eq:conductivity}) given above, we can calculate \textit{ab initio} the conductivity tensor, from which we can easily obtain the dielectric tensor.
Further, the $n^\pm$ refractive indices are given by \cite{oppeneer01}
\begin{equation}
\label{eq:refr}
(n^{\pm})^2 = \epsilon_{xx}\pm i\epsilon_{xy}\sin\phi_t ,
\end{equation}
which, making a Taylor expansion for $\epsilon_{xy}\ll \epsilon_{xx}$, gives
\begin{equation}
\label{eq:qdef}
n^+-n^- \simeq \bar{n}Q\sin\phi_t .
\end{equation}
The angles of incidence and transmittance are related through Snell's law,
\begin{equation}
\label{eq:sn}
\bar{n} \sin\phi_t = {\sin\phi_i},
\end{equation}
where $n_0 =1 $ for the refractive index of vacuum.
Substitution of the above expressions in Eq.\ (\ref{eq:kerr}) gives
\begin{small}
	\begin{eqnarray}
	\label{eq:LMOKE}
	\! \! \! \! \! \! \theta_{\rm K}^{L}+i\varepsilon_{\rm K}^{L}&=&-\frac{\epsilon_{xy}\sin 2\phi_i}{2(1-\epsilon_{xx})[\cos \phi_i (\epsilon_{xx}-\sin^2 \phi_i)+\sin^2\phi_i(\epsilon_{xx}-\sin^2 \phi_i)^{1/2}]} \quad \text{(s-polarization),}\\
	\! \! \! \! \! \! \theta_{\rm K}^{L}+i\varepsilon_{\rm K}^{L}&=&-\frac{\epsilon_{xy}\sin 2\phi_i}{2(1-\epsilon_{xx})[\cos \phi_i (\epsilon_{xx}-\sin^2 \phi_i)-\sin^2\phi_i(\epsilon_{xx}-\sin^2 \phi_i)^{1/2}]} \quad \text{(p-polarization).}
	\end{eqnarray}
\end{small}

Due to the dependence  of the Kerr rotation angle on $\epsilon_{xx}$ and consequently on $\sigma_{xx}$, a good choice of the Drude parameters plays a role for the determination of the theoretical Kerr angle. As mentioned before,
we prefer to use values of the Drude peak extracted from the average of the measured DC resistivity $\rho$, given in Table \ref{resistance}, and for the Drude lifetime we use the values from Ref.\ \cite{ordal85}.
{To obtain the bare Drude conductivity $\sigma_{_{\rm D}}$ we therefore use
	\begin{equation}
	\sigma_{_{\rm D}}=\rho^{-1} - {\rm Re}[ \sigma_{xx}^{inter}(0)],
	\end{equation}
	where $\sigma_{xx}^{inter}(0)$ is the \textit{ab initio} calculated interband contribution to the DC conductivity,
	shown in Fig.\ \ref{fig:diag}. The calculated values of the Drude conductivity are $\sigma_{_{\rm D}} (\rm Pt) =4.6824\times10^{-2}$ $(\mu\Omega$cm$)^{-1}$ and $\sigma_{_{\rm D}} (W)=4.8\times10^{-3}$ $(\mu\Omega$cm$)^{-1}$ for Pt and W, respectively.}

\begin{figure}[ht!]
	\includegraphics[width=10cm]{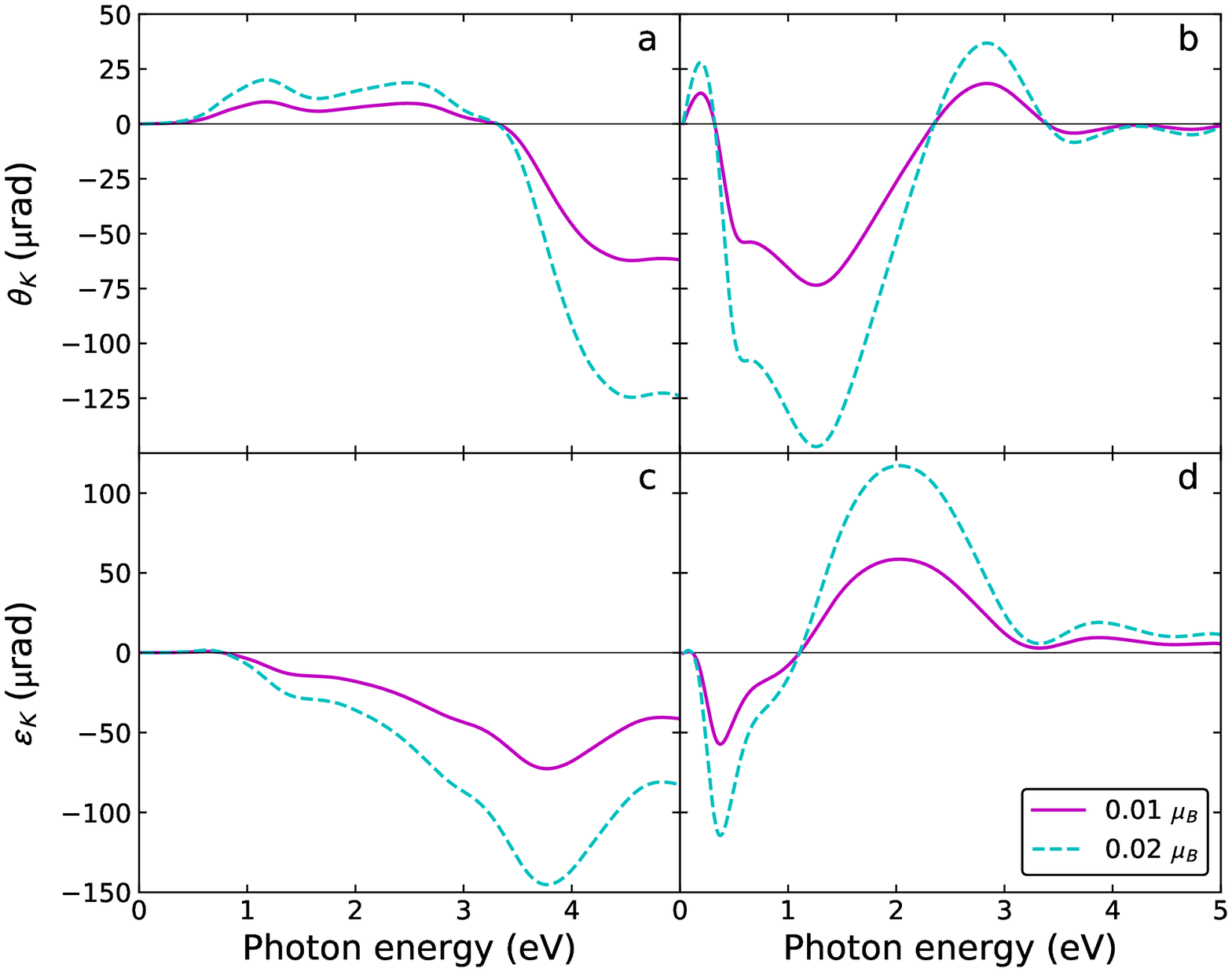}
	\caption{\textit{Ab initio} calculated Kerr rotation $\theta_{\rm K}$ and ellipticity $\varepsilon_{\rm K}$ for the longitudinal configuration with s-polarized incident light. Results are given for two different induced magnetizations, $M_{ind}^{ref}=0.01$ $\mu_{\rm B}$ (red lines) and  $M_{ind}^{ref}=0.02$ $\mu_{\rm B}$ per atom (blue dashed lines), in the left panels for Pt and in the right panels for W.
		For both metals we show the Kerr effect considering a Drude peak defined by the average resistivity of the thin film samples.}
	\label{fig:MOKE}
\end{figure}

Having predicted the surface magnetization of the sample created by the SHE through Eqs.\ (\ref{eq:spin_pot_sm}) and (\ref{eq:mag}) we can now proceed to calculate the expected Kerr rotation in an identical set-up as in the experiment.
The experimental measurement were performed with:
1) a photon energy $\hbar \omega=2.414$ eV of the probing laser, and 2)
an angle of incidence of $\phi_i=37^{\circ}$.
We calculate the longitudinal Kerr rotation angle for a magnetized sample using Eq.\ (\ref{eq:LMOKE}) above. The Kerr angles for nonmagnetic Pt and W in their ground state are of course zero.
In order to provide a reference, we introduce a fictitious static magnetic field in the calculations, through which we induce a magnetization of $M_{ind}^{ref}=0.01\,\mu_{\rm B}$ per atom in both Pt and W, and we calculate the nonzero longitudinal Kerr rotation for this magnetized state.
The longitudinal Kerr rotation calculated as a function of the photon energy for a magnetization $M_{ind}^{ref}=0.01\, \mu_{\rm B}$ per atom is shown for Pt and W in Fig.\ \ref{fig:MOKE} together with the calculated Kerr angle for a twice as large magnetization (obtained by increasing the fictitious magnetic field). This calculation shows, as one would expect, that the Kerr angle at each photon energy increases linearly with the small reference  magnetization. We note also that the calculated $\theta_{\rm K}$ at the experimental photon energy (2.414 eV) is positive for both Pt and W because we assume a positive $M_{ind}^{ref}$ in either case. However, the sign of the SHE is opposite in the two metals, which implies that the spin accumulation and the Kerr rotation in W should be negative, as observed in the experiment.

We can now calculate the proportionality constant between the complex Kerr effect and the induced magnetization as the ratio
\begin{equation}
\label{eq:prop}
\alpha=\frac{\Phi_{\rm K}^{L,bulk}}{M_{ind}^{ref}}.
\end{equation}
For Pt we obtain $\alpha =
(0.925 - 2.618 i) \cdot 10^{-3}$ rad/$\mu_{\rm B}$ at a photon energy $\hbar\omega$ of 2.414 eV.

\section{Depth sensitivity of the longitudinal MOKE}
For a more accurate calculation of the expected Kerr effect in the experiment we need to consider the depth sensitivity of MOKE \cite{Traeger1992sm} and that the spin accumulation is not constant over the the thickness of the film.
If we consider a material uniformly magnetized, each layer $[z,z+dz]$ of the material will provide a Kerr effect contribution:
\begin{equation}
\label{eq:diff_rot}
d\Phi_{\rm K}^L(z)=\Phi_{\rm K}^{L,bulk}(z)\frac{4\pi i \bar{n}}{\lambda}\mathrm{e}^{-\frac{4\pi i \bar{n} z}{\lambda}}dz,
\end{equation}
where $\lambda$ is the wavelength of the probing laser.
Combining this equation with Eqs.\ (\ref{eq:spin_pot_sm}) and (\ref{eq:mag}) we can calculate the expected complex Kerr rotation for a film with thickness $t$ as
\begin{equation}
\label{eq:exp_Kerr}
\Phi_{\rm K}^{L,tot}=\int_0^t M^y(z)\,\alpha \, \frac{4\pi i\bar{n}}{\lambda}\mathrm{e}^{-\frac{4\pi i \bar{n}z}{\lambda}}dz.
\end{equation}
The complex refractive index $\bar{n} = (n^+ - n^-)/2$ can be easily calculated;
we obtain $\bar{n}=2.12-3.79i$ for bulk Pt at 2.414 eV photon energy.

As a next step, we can analytically integrate Eq.\ (\ref{eq:exp_Kerr}) which provides the following expression for the complex longitudinal Kerr effect ($\theta_{\rm K}^m +i \varepsilon_{\rm K}^m = \Phi_{\rm K}^{L,tot}$) that is expected in a measurement,
\begin{equation}
\label{eq:meask_sm}
\theta_{\rm K}^m+i\varepsilon_{\rm K}^m=\frac{\sdl\theta_{\rm SH}\,\rho j_xD(E_{\rm F})F\Phi_{\rm K}^{L,bulk}\mathrm{e}^{\frac{t}{2\sdl}}}{2\cosh(\frac{t}{2\sdl})}\kappa\left(\frac{(\mathrm{e}^{-\kappa^-t}-1)\mathrm{e}^{-\frac{t}{\sdl}}}{\kappa^-}-\frac{\mathrm{e}^{-\kappa^+t}-1}{\kappa^+}\right),
\end{equation}
where we have defined $\kappa= (4\pi i \bar{n} \cos\psi )/{\lambda}$, $\cos \psi= (1-\sin^2\phi_i/\bar{n}^2 )^{1/2}$, and $\kappa^{\pm}=\kappa \pm 1/\sdl$.
All quantities in Eq.\ (\ref{eq:meask_sm}) are obtained from our \textit{ab initio} calculations or defined from experiment ($t$, $j_x$) except the spin diffusion length \sdl. We can thus use Eq.\ (\ref{eq:meask_sm}) to calculate the expected Kerr rotation as a function of film thickness for a given spin diffusion length. We find that the theoretical spin diffusion length that describes
best the experimental data is $\sdl = 8.3$~nm, which compares well with that obtained from an unconstrained fit of $\theta_{\rm K}$, as shown in Fig.~5 of the main text.

Lastly, we mention that one can take the real part on both sides of Eq.\ (\ref{eq:meask_sm}) and rewrite it to obtain the spin Hall angle $\theta_{\rm SH}$ as a function of the other above-mentioned quantities,
\begin{equation}
\theta_{\rm SH}=\frac{\theta_{\rm K}^m}{j_x}\frac{2\cosh(\frac{t}{\sdl})}{\sdl\rho j_xD(E_{\rm F})F\, } \, {\rm Re} \left[ \frac{1}{\Phi_{\rm K}^{L,bulk}} \frac{\kappa^-\kappa^+}{ \kappa[\kappa^+(\mathrm{e}^{-\kappa^-t}-1)\mathrm{e}^{-\frac{t}{\sdl}}-\kappa^-(\mathrm{e}^{-\kappa^+t}-1)]}\right] .
\end{equation}

\section{Fits of the Kerr rotation as a function of platinum thickness}

As shown above, MOKE requires only knowledge of the dielectric tensor of Pt in order to provide a quantitative estimate of the average current-induced spin accumulation $M^y$. Extracting the spin diffusion parameters from $\theta_{\rm K}$, however, necessarily requires a model of the spin accumulation as a function of thickness. This is an issue common to all experimental probes of the SHE, which is partly responsible for the large parameter spread reported in the literature. Because of its simplicity, the large majority of experimental studies adopts the one-dimensional drift-diffusion approach, which models the spin accumulation in the direction perpendicular to the charge current by assuming thickness-independent parameters, namely $\rho$, $\sigma_{xz}^{\rm SH}$, and $\sdl$ \cite{Zhang2000sm,Chen2013}. Recently, however, theoretical \cite{Liu2014} and experimental studies \cite{Rojas2014sm,Nguyen2016sm,Sagasta2016sm} have pointed out the need to consider the dependence of $\sdl$ on $\rho$. An inverse proportionality relationship $\sdl \propto \rho^{-1}$ is expected if the Elliott-Yafet mechanism dominates spin relaxation in Pt. In such a case, the relevant parameter to compare between different experiments is the product $\sdl \rho$, which has been reported to vary between 0.6 and 1.3~f$\Omega$m$^2$ in recent work \cite{Niimi2013sm,Nguyen2016sm,Sagasta2016sm}. The fit of $\theta_{\rm K}$ reported in Fig.~5 of the main text performed at constant $\rho= 20.6$~$\mu\Omega$cm (the average resistivity of our films, measured at high current, see Table~S1) gives $\sdl = 11.4$~nm, hence $\sdl\rho = 2.3$~f$\Omega$m$^2$. The fit performed by interpolating the experimental resistivity and keeping the product $\sdl \rho$ as free parameter (dashed line in Fig.~5) gives $\sdl\rho = 2.6$~f$\Omega$m$^2$. With this hypothesis, \sdl varies from about 10~nm in Pt(5~nm) to 16~nm in Pt(60~nm). Both the $\sdl \rho$ values extracted from these fits are considerably larger than $\sdl\rho = 0.8$~f$\Omega$m$^2$ reported for spin-orbit torque measurements of Pt/Co bilayers as well as relative to $\sdl\rho = 1.3$~f$\Omega$m$^2$ \cite{Niimi2013sm} and $\sdl\rho = 0.6$~f$\Omega$m$^2$ \cite{Sagasta2016sm} reported for spin absorption measurements using nonlocal spin valves. The discrepancy with respect to spin torque measurements can possibly be explained by the presence of spin-dependent scattering and proximity effects in ferromagnetic/nonmagnetic metal bilayers, which are absent in our samples. Differences with respect to the nonlocal spin valve measurements, in which Pt is not in direct contact with a ferromagnet, may be due to the different assumptions required to extract $\sdl$ from MOKE and spin absorption experiments. Whereas MOKE requires knowledge of the optical constants of the probed material, lateral spin valve measurements require  modelling of three-dimensional spin diffusion processes in the electrodes as well as the knowledge of additional variables, such as the spin diffusion length of the light metal channel and the spin resistances of the light metal/heavy metal interface, which are usually assumed to be transparent. Other differences may arise because of temperature, current density, and sample quality.

\begin{figure}[t]
	\includegraphics[width=14cm]{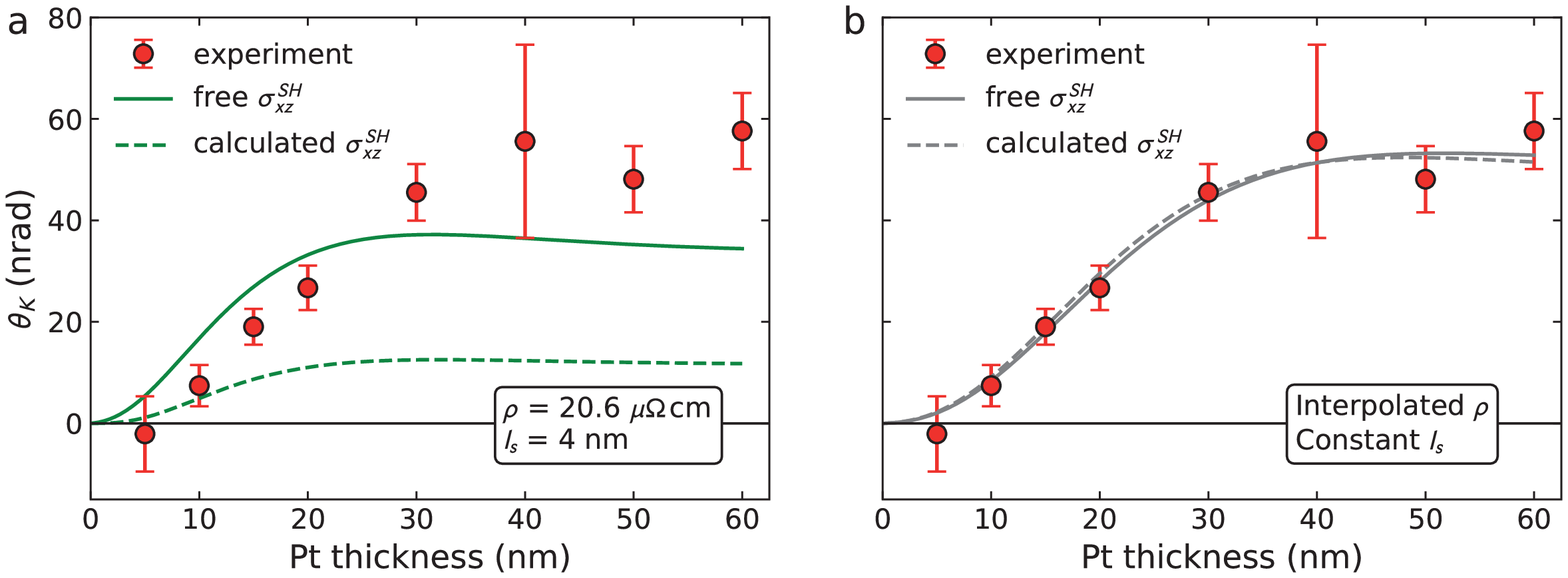}
	\caption{Comparison between the Kerr rotation calculated using Eq.~(2) of the main text and experimental data. The calculations assume (a) average resistivity $\rho = 20.6 \, \mu\Omega cm$ and $\sdl = 4$~nm and (b) interpolated resistivity and constant $\sdl$. The solid curves fit the spin Hall conductivity $\sigma_{xz}^{\rm SH}$, whereas the dashed curves are obtained taking the theoretical value of $\sigma_{xz}^{\rm SH} = 1890\, \Omega^{-1} {\rm cm}^{-1}$.}
	\label{fig:FITS}
\end{figure}
In order to test the validity of our conclusions, we have performed additional fits of $\theta_{\rm K}$ as a function of Pt thickness using different assumptions. Figure~\ref{fig:FITS}
(a) shows $\theta_{\rm K}$ calculated by taking a constant resistivity $\rho = 20.6$~$\mu\Omega$cm and $\sdl = 4$~nm, as expected by assuming $\sdl\rho = 0.8$~f$\Omega$m$^2$, similar to Refs.~\cite{Rojas2014sm,Nguyen2016sm,Sagasta2016sm}. The dashed line is a calculation with no free parameters using the \textit{ab initio} value of the spin Hall conductivity $\sigma_{xz}^{\rm SH}= 1890$ $\Omega^{-1}{\rm cm}^{-1}$. The solid line is a fit of $\theta_{\rm K}$ with $\sigma_{xz}^{\rm SH}= 5740$ $\Omega^{-1}{\rm cm}^{-1}$. The poor quality of the fits shows that, within our model, values of \sdl that are significantly shorter than 10~nm do not capture the thickness dependence of $\theta_{\rm K}$. Also, fits of $\theta_{\rm K}$ that consider the thickness-dependent $\rho$ and a constant \sdl, shown in  Fig.~\ref{fig:FITS} (b), give $\sdl = 13$~nm for the theoretical $\sigma_{xz}^{\rm SH}= 1890$ $\Omega^{-1}{\rm cm}^{-1}$ (dashed line) and $\sdl = 14.4$~nm for $\sigma_{xz}^{\rm SH}= 1730$ $\Omega^{-1}{\rm cm}^{-1}$ (solid line).

Before concluding this Section, we caution that some assumptions in the analysis of thickness-dependent spin accumulation measurements should be tested further. First, the solution of the spin diffusion equations \cite{Zhang2000sm,Chen2013}, namely the spin accumulation profile along $z$ (Eq.~(1) of the main text), is derived by assuming constant and uniform $\rho$ and $\sigma_{xz}^{\rm SH}$ throughout the nonmagnetic metal. This is very often not the case inside thin films, where $\rho$ varies from one interface to the other. Because of nonspecular scattering at the interfaces, the variation of $\rho$ inside a thin film is not the same as the variation of $\rho$ measured in films of different thickness. A related problem is that drift-diffusion theory assumes that \sdl is much larger than the electron mean free path, which is often not the case in thin films. Second, different mechanisms besides Elliott-Yafet scattering can contribute to spin relaxation, also in Pt \cite{Ryu2016}. Third, recent \textit{ab initio} theoretical calculations predict a significant enhancement of $\sigma_{xz}^{\rm SH}$ near ferromagnetic interfaces \cite{Wang2016sm,Freimuth2015}. Such an enhancement may be present also at a metal-vacuum interface, although it has not been predicted yet.
These considerations show that refined models of the spin accumulation are required in order to properly account for thickness-dependent effects. Nonetheless, our data show that the saturation of $\theta_{\rm K}$ in Pt occurs over a length scale of about 10~nm, independently of the model used for fitting.%

\end{document}